\documentclass[aip,apl,reprint,a4paper,floatfix,amsmath,amssymb,amsfonts,noshowpacs,citeautoscript]{revtex4-2}
\usepackage{newtxtext,newtxmath}
\usepackage[T1]{fontenc}
\usepackage{graphicx}
\usepackage[dvipsnames]{xcolor}
\usepackage{soul} 
\usepackage{floatrow}
\usepackage[
,textwidth=17.5cm
,textheight=23.5cm
,verbose
,pdftex
]{geometry}
\usepackage{xspace}

\usepackage[pdftex]{hyperref}
\hypersetup{
  pdftitle={Dualtronics},
	colorlinks,
	citecolor=blue,
	linkcolor=blue
	}

\makeatletter
\newcommand*{\refig}[2]{\hyperref[#1]{\ref*{#1}(#2)}}
\makeatother

\DeclareMathAlphabet{\mathsf}{OT1}{\sfdefault}{m}{n}
\SetMathAlphabet{\mathsf}{bold}{OT1}{\sfdefault}{b}{n}

\begin{document}
\graphicspath{{./figures/}}
\title{Epitaxial lattice-matched Al$_{0.89}$Sc$_{0.11}$N/GaN distributed Bragg reflectors}

\author{L.~van~Deurzen}
\email[Electronic mail: ]{lhv9@cornell.edu}
\affiliation{School of Applied and Engineering Physics, Cornell University, Ithaca, New York 14853, USA}
\author{T.-S.~Nguyen}
\email[Electronic mail: ]{tn354@cornell.edu}
\affiliation{Department of Materials Science and Engineering, Cornell University, Ithaca, New York 14853, USA}
\author{J.~Casamento}
\affiliation{Department of Materials Science and Engineering, Cornell University, Ithaca, New York 14853, USA}
\author{H.G.~Xing}
\affiliation{Department of Materials Science and Engineering, Cornell University, Ithaca, New York 14853, USA}
\affiliation{Department of Electrical and Computer Engineering, Cornell University, Ithaca, New York 14853, USA}
\affiliation{Kavli Institute at Cornell for Nanoscale Science, Cornell University, Ithaca, New York 14853, USA}
\author{D.~Jena}
\affiliation{School of Applied and Engineering Physics, Cornell University, Ithaca, New York 14853, USA}
\affiliation{Department of Materials Science and Engineering, Cornell University, Ithaca, New York 14853, USA}
\affiliation{Department of Electrical and Computer Engineering, Cornell University, Ithaca, New York 14853, USA}
\affiliation{Kavli Institute at Cornell for Nanoscale Science, Cornell University, Ithaca, New York 14853, USA}

\begin{abstract}
We demonstrate epitaxial lattice-matched Al$_{0.89}$Sc$_{0.11}$N/GaN ten and twenty period distributed Bragg reflectors (DBRs) grown on c-plane bulk n-type GaN substrates by plasma-enhanced molecular beam epitaxy (PA-MBE). Resulting from a rapid increase of in-plane lattice coefficient as scandium is incorporated into AlScN, we measure a lattice-matched condition to $c$-plane GaN for a Sc content of just 11\%, resulting in a large refractive index mismatch $\mathrm{\Delta n}$ greater than 0.3 corresponding to an index contrast of $\mathrm{\Delta n/n_{GaN}}$ = 0.12 with GaN. The DBRs demonstrated here are designed for a peak reflectivity at a wavelength of 400 nm reaching a reflectivity of 0.98 for twenty periods. It is highlighted that AlScN/GaN multilayers require fewer periods for a desired reflectivity than other lattice-matched Bragg reflectors such as those based on AlInN/GaN multilayers.
\end{abstract}

\maketitle

In recent years, the ultra-wide bandgap material Al$_{1-x}$Sc$_{x}$N has emerged as a subject of intense research interest, particularly within the domain of semiconductor electronics. The intentional introduction of scandium into AlN has been found to result in a remarkable enhancement of its piezoelectric coefficients \cite{akiyamaEnhancementPiezoelectricResponse2009, tasnadiOriginAnomalousPiezoelectric2010} and, notably, the emergence of ferroelectricity \cite{fichtnerAlScNIIIVSemiconductor2019}. More recently, perhaps not surprisingly due to the anisocrystalline alloying of rocksalt ScN and wurtzite AlN, it is also being realised that certain optical properties of wurtzite AlScN are improved upon over those of AlN, including the second order non-linear optical coefficients \cite{yoshiokaStronglyEnhancedSecondorder2021}. Moreover, the possibility of lattice matching to GaN, along with the recent developments of their thin film synthesis, including reactive sputtering and epitaxy, predict a positive outlook on the integration of wurtzite AlScN with existing group III/N semiconductor technologies.

While significant efforts have been dedicated to study piezoelectricity and ferroelectricity of AlScN for integration in RF \cite{greenScAlNGaNHighElectronMobility2019, krauseAlScNGaNHEMTs2023,casamentoFerroHEMTsHighCurrentHighSpeed2022} and memory applications \cite{wangEpitaxialFerroelectricScAlN2022}, reports on optoelectronic applications of AlScN remain scarce. In spite of this, the existing properties that make AlScN an attractive material for GaN-based electronics could also make it a promising material for optoelectronic integration. For example, lattice-matched AlScN is a good barrier for GaN high-electron mobility transistors (HEMTs) to overcome the critical thickness limitation in (Al,Ga)N/GaN HEMTs \cite{kaziorHighPowerDensity2019}. Similarly, lattice-matched AlScN can be competitive to replace or outperform AlN, AlGaN, and AlInN in optoelectronic applications where minimizing crystal degradation and crack formation are of high interest. 


\begin{figure*}[t]
\includegraphics[width=\textwidth]{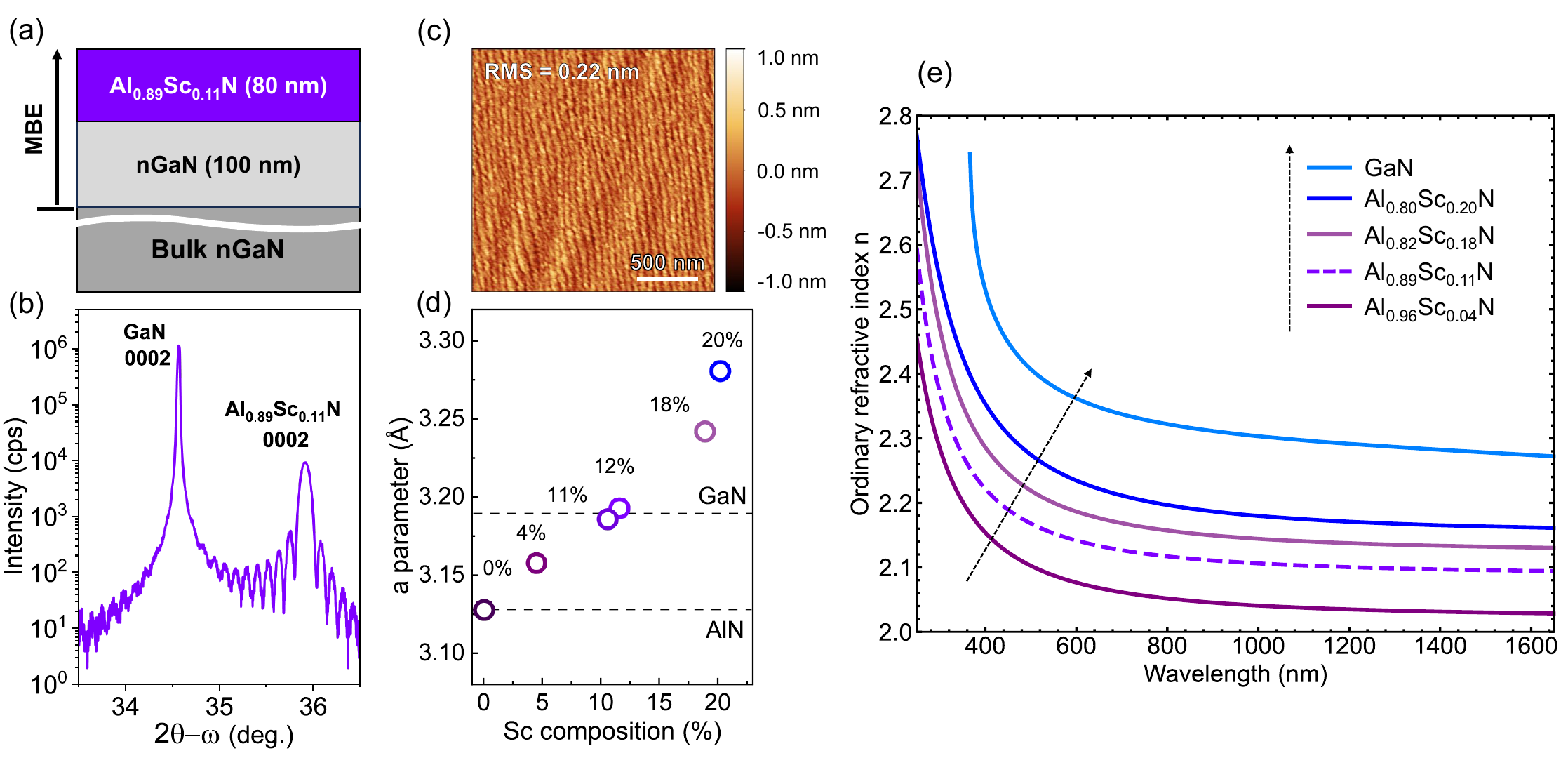}
\caption{(a) Structure of a typical AlScN film grown on a bulk n$^{+}$GaN substrate for lattice parameters and refractive index dispersion characterizations. (b) Symmetric 2$\theta$-$\omega$ XRD scan of thin Al$_{0.89}$Sc$_{0.11}$ on bulk n-type GaN. (c) 2x2 $\mu$m$^2$ AFM micrograph shows clear atomic steps with rms = 0.22 nm. (d) AlScN in-plane lattice parameter vs. Sc composition measured from reciprocal space mapping of the AlScN (10$\bar{1}$5) peak. (e) Ordinary refractive index n vs. vacuum wavelength for Al$_{1-x}$Sc$_{x}$N films of various Sc content, as well as GaN; the dashed line corresponds to the nominally lattice-matched Al$_{0.89}$Sc$_{0.11}$N sample. \label{fig:Figure_1}}
\end{figure*}


One particular example is in GaN-based distributed Bragg reflectors (DBRs) for nitride-based microcavities and vertical-cavity surface-emitting lasers (VCSELs) and resonant cavity light-emitting diodes (RCLEDs) \cite{takeuchiGaNbasedVerticalcavitySurfaceemitting2018,carlinCrackfreeFullyEpitaxial2005,inabaGaNEuOBased2020}. AlN/GaN and AlGaN/GaN are the first semiconductor-based epitaxial DBRs explored for GaN-based VCSELs \cite{khanReflectiveFiltersBased1991,ngDistributedBraggReflectors1999,waldripStressEngineeringMetalorganic2001}. Unlike dielectric DBRs (e.g. SiO$_2$-based), Al(Ga)N-based epitaxial DBRs do not require complex fabrication techniques like lift-off and bonding\cite{takeuchiGaNbasedVerticalcavitySurfaceemitting2018}. However, growing thick, high-quality Al(Ga)N/GaN DBRs remains challenging and requires complex strain engineering schemes due to the large lattice mismatch between AlN and GaN \cite{waldripStressEngineeringMetalorganic2001,huangCrackfreeGaNAlN2006}. Low refractive index porous GaN can also induce a significant refractive index mismatch for GaN-based DBRs, but is limited by its complicated etching process and its potential for degrading structural integrity \cite{takeuchiGaNbasedVerticalcavitySurfaceemitting2018}.

Lattice-matched Al$_{0.82}$In$_{0.18}$N/GaN is a promising alternative to circumvent degradation of crystal and optical properties due to strain relaxation induced by lattice mismatch \cite{carlinHighqualityAlInNHigh2003, carlinCrackfreeFullyEpitaxial2005, gacevicInAlNGaNBragg2010}. However, the synthesis of high-quality AlInN thin films and AlInN/GaN layers is difficult because of the large difference in optimal growth temperatures for InN and AlN \cite{bergerGrowthAlInNGaN2015,gacevicHighQualityInAlN2011,chungGrowthStudyImpurity2011}. Furthermore, the refractive index mismatch $\Delta$n $\approx$ 0.2 between Al$_{0.82}$In$_{0.18}$N and GaN is quite low (relative contrast $\mathrm{\Delta n/n_{GaN}}$ $\approx$ 0.08), meaning that more AlInN/GaN pairs are needed to achieve the same reflectivity demonstrated by Al(Ga)N/GaN DBRs \cite{butteRecentProgressGrowth2005, takeuchiGaNbasedVerticalcavitySurfaceemitting2018}. Lastly, GaN-based DBRs are often epitaxially integrated as the bottom reflector in photonic devices, so a high crystal quality is critical to achieving any further epitaxial integration of active layers.  

Replacing AlInN with AlScN lattice-matched to GaN can address these limitations since AlScN growth conditions are more compatible with GaN \cite{hardyEpitaxialScAlNEtchStop2017a,casamentoStructuralPiezoelectricProperties2020}, and the lattice-matched condition occurs at a higher Al composition \cite{dinhLatticeParametersScxAl12023,dengBandgapAl1XScxN2013}, which could yield a higher index mismatch with GaN. To this end, it is essential to determine the AlScN/GaN lattice-matched condition and the dependence of the refractive index on Al$_{1-x}$Sc$_{x}$N alloy composition, $x$.

In this work, by spectroscopic ellipsometry, we detail the dispersion of the refractive index of thin films of Al$_{1-x}$Sc$_{x}$N near its lattice-matched condition with GaN. By studying films of approximately 80-100~nm thickness grown on bulk $c$-plane metal-polar n$^{+}$GaN by PA-MBE, we find that the lattice-matched condition occurs at a scandium content of $x$~=~0.11. At this composition, we infer a refractive index mismatch $\Delta$n of approximately 0.3 and index contrast $\mathrm{\Delta n/n_{GaN}}$ = 0.12 with respect to GaN, throughout the UV-A, visible, and near-infrared (NIR) spectral regimes. With this significant index mismatch and low optical losses due to the ultra-wide bandgap of approximately 5.6~eV for Al$_{0.89}$Sc$_{0.11}$N \cite{casamentoEpitaxialScxAl1XN2022, wangMolecularBeamEpitaxy2020}, high-reflectivity distributed Bragg reflectors are feasible for wavelengths limited by the bandgap of GaN (3.4 eV). In comparison with AlInN/GaN Bragg reflectors, the higher index mismatch in AlScN/GaN near the lattice matched scandium composition predicts the need for fewer periods and, therefore, a lower total film thickness for a given desired peak reflectivity.
Furthermore, we experimentally demonstrate such Bragg reflectors with peak reflectivity at a wavelength of 400~nm by growing ten- and twenty period multilayers, yielding a peak reflectivity of 0.98.

\begin{figure*}[t]
\includegraphics[width=\textwidth]{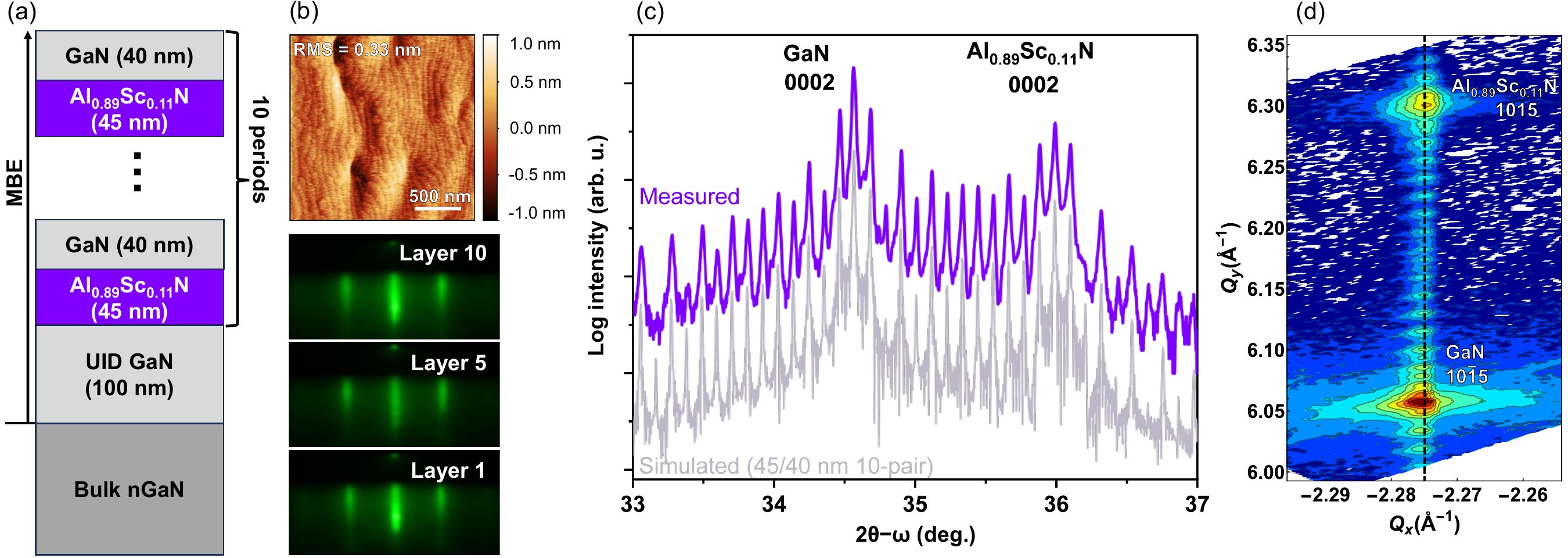}
\caption{(a) Schematic of the ten period Al$_{0.89}$Sc$_{0.11}$N/GaN distributed Bragg reflector on bulk n$^{+}$GaN. (b) AFM micrograph with clear atomic steps (rms~=~0.34 nm) and streaky RHEED evolution of AlScN layers along the <110> zone axis. (c) Symmetric 2$\theta$-$\omega$ scan shows a strong AlScN (0002) peak; the AlScN layer peak and Pendell\"{o}sung fringes are well-described by the simulated 10-pair AlScN/GaN 45/40~nm diffraction pattern. (d) XRD reciprocal space map of (10$\bar{1}$5) reflections confirm that the AlScN and GaN layers are pseudomorphically grown on the bulk n$^{+}$GaN substrate. \label{fig:Figure_2}}
\end{figure*}


All AlScN films and multilayer structures in this work were grown on the c-plane of bulk silicon-doped Ga-polar n-type GaN (n$^{+}$GaN) from Ammono. A Veeco GenXplor MBE reactor was used for all growths in this study. Scandium, aluminium, gallium, and silicon were provided using effusion K-cells. Active nitrogen species was provided using an RF plasma source with a 1.95 sccm nitrogen flowrate and 200~W RF power. The surface morphology was characterized by an Asylum Research Cypher ES atomic force microscope (AFM). A PANalytical Empyrean system with Cu K$_{\alpha1}$ radiation was used for X-ray diffraction (XRD), X-ray reflectivity (XRR) and reciprocal space mapping (RSM) to determine crystal structure, out-of-plane and in-plane lattice constants, and thin film thickness, respectively. The scandium composition was measured by energy-dispersive X-ray spectroscopy using a Zeiss LEO 1550 FESEM equipped with a Bruker energy dispersive X-ray spectroscopy (EDS) silicon drift detector (SDD). The refractive index and optical loss dispersion for in-plane polarized light were measured with a Woollam RC2 spectroscopic ellipsometer using the single layers of AlScN grown on the bulk n$^{+}$GaN substrates. Reliable measurements were enabled by applying Mueller matrix ellipsometry in an optical window ranging from 193-1690~nm. Finally, the reflectivity spectra of the AlScN/GaN DBRs were then measured by using an Agilent Cary 5000 UV-Vis-NIR spectrophotometer. The spectra were calibrated by using a UV-enhanced Aluminum mirror with well-known reflectivity.

The AlScN films were grown under nitrogen-rich conditions to promote scandium incorporation and preserve the wurtzite phase purity \cite{hardyEpitaxialScAlNEtchStop2017a,casamentoStructuralPiezoelectricProperties2020}. A metal (Sc+Al) to nitrogen (III/V) ratio of 0.7 was employed. GaN was grown under metal-rich conditions to promote the step-flow growth mode for high crystallinity. The growth rates for AlScN and GaN were 3.0~nm/min and 3.8~nm/min, respectively. All growths were monitored in situ by reflection high-energy electron diffraction (RHEED).

Thin films of AlScN were grown on bulk n$^{+}$GaN substrates to determine the lattice-matched condition. For each sample, a 100~nm Si-doped n$^{+}$GaN layer was grown at a substrate temperature of 630~$^{\circ}$C measured by a thermocouple. Excess Ga was fully consumed before the N-rich AlScN growth, followed by 80-100~nm AlScN grown at 530~$^{\circ}$C thermocouple temperature. Note that the thermocouple substrate temperatures are approximately 50~$^{\circ}$C below the true temperature. Based on the measured ordinary refractive index and lattice matched composition, Al$_{1-x}$Sc$_{x}$N and GaN quarter-wavelength thicknesses were calculated for a peak reflectivity targeted at a vacuum wavelength of 400~nm. After the growth of a 100~nm unintentionally doped GaN buffer layer at 530~$^{\circ}$C, ten period and twenty period DBR structures were grown with the targeted quarter wavelength layer thicknesses. GaN was grown at the optimal growth temperature of AlScN to prevent growth interruption between layers. The Ga flux was calibrated to achieve approximately 8 seconds of Ga droplets for a 10 minute GaN growth; excess Ga consumption was accounted for in the total growth time to control the GaN thickness precisely. Similar AlScN layer thicknesses were achieved by using the same III/V ratio (0.7) and growth time.

The thin (80-100~nm) AlScN films, as depicted in Fig.~\ref{fig:Figure_1}(a), are stabilized in the wurtzite phase, as confirmed by the strong (0002) AlScN diffraction peak near 2$\theta$ = 36$^{\circ}$ [Fig.~\ref{fig:Figure_1}(b)] \cite{hardyEpitaxialScAlNEtchStop2017a,casamentoStructuralPiezoelectricProperties2020}. For AlScN films with Sc composition around 12\%, strong Pendell\"{o}sung fringes are observed in the symmetric XRD scan, suggesting high interface quality between AlScN and n$^{+}$GaN layers. More importantly, when AlScN is nearly fully strained to the GaN substrate at a scandium incorporation of 11\%, the two-dimensional step-flow growth mode and surface root-mean-square (rms) roughness below 3~$\Angstrom$ could be achieved despite the nitrogen-rich growth condition [Fig.~\ref{fig:Figure_1}(c)]. From in-plane lattice constants at different Sc content [Fig. \ref{fig:Figure_1}(d)], the lattice-matched condition is determined to lie between 11\% and 12\%~Sc. These results suggest that nominally lattice-matched AlScN films of 80-100~nm thickness can be well integrated with GaN to achieve pseudomorphic AlScN/GaN multilayer structures which are highly crystalline and display sharp interfaces.

 Figure~\ref{fig:Figure_1}(e) shows the ordinary refractive index of AlScN as a function of Sc composition and wavelength determined from Mueller matrix ellipsometry measurements. The refractive index difference between Al$_{0.89}$Sc$_{0.11}$N and GaN is $\Delta$n~=~0.3 ($\mathrm{\Delta n/n_{GaN}}$ = 0.12) for a vacuum wavelength $\lambda$~=~400~nm; this is significantly larger than $\Delta$n~$\approx$~0.2 ($\mathrm{\Delta n/n_{GaN}}$~=~0.06-0.08) for lattice-matched AlInN/GaN.\cite{gacevicInAlNGaNBragg2010, butteRecentProgressGrowth2005} The larger index mismatch is enabled partly by AlScN having a larger lattice-matched aluminum composition than AlInN \cite{dinhLatticeParametersScxAl12023, dengBandgapAl1XScxN2013} due to the rapid increase of in-plane lattice constant with Sc composition [Fig.~\ref{fig:Figure_1}(d)]. It is important to note that various lattice-matched compositions between 9\% and 18\%~Sc have been reported for AlScN grown by different methods and conditions \cite{dinhLatticeParametersScxAl12023,wangMolecularBeamEpitaxy2020,casamentoStructuralPiezoelectricProperties2020,dzubaEliminationRemnantPhases2022}. Therefore, the specific design parameters (refractive index, layer thickness, lattice matched condition) would vary depending on the growth conditions and specific structural and optical properties of AlScN films. 

Accurate thickness control is also critical for a good DBR since the reflectivity depends strongly on the layer thicknesses. Figure \ref{fig:Figure_2}(a) shows the AlScN/GaN DBR multilayer structure designed with intended thicknesses of 45~nm and 40~nm for Al$_{0.89}$Sc$_{0.11}$N and GaN, respectively, for $\lambda$~=~400 nm  ($\lambda$/4n for each layer). To precisely control layer thicknesses, the molecular beam fluxes and substrate temperature were kept constant throughout the growth. The substrate thermocouple temperature (530~$^{\circ}$C) is lower than the optimal growth temperature of GaN but is optimal for AlScN \cite{hardyEpitaxialScAlNEtchStop2017a,dzubaEliminationRemnantPhases2022} and helps minimize the growth interruption between alternating AlScN and GaN layers. The growth conditions reported here are more easily controlled than in AlInN/GaN multilayer growths, which requires careful temperature and flux control due to high In desorption and InN decomposition rates at temperatures suitable for GaN and AlN growths \cite{gacevicInAlNGaNBragg2010}. Figure~\ref{fig:Figure_2}(b) shows a streaky RHEED pattern along the <110> zone axis in all AlScN layers grown under nitrogen-rich conditions. This is in accordance with the RHEED pattern and surface morphology in the nominally lattice-matched single-layer AlScN heterostructure [Fig.~\ref{fig:Figure_1}(c)] and in other studies \cite{casamentoStructuralPiezoelectricProperties2020,dzubaEliminationRemnantPhases2022,dinhLatticeParametersScxAl12023}. Furthermore, the smooth surface morphology is maintained after ten [Fig.~\ref{fig:Figure_2}(b)] and even twenty [see Supplementary Material] Al$_{0.89}$Sc$_{0.11}$/GaN periods. Specifically, an rms surface roughness of 0.33~$\Angstrom$ and clear atomic steps were achieved for a total growth thickness of 950~nm in the ten period DBR sample despite the nitrogen rich AlScN growth, highlighting the high crystallinity and interface qualities of nominally lattice-matched growth conditions.

\begin{figure}[t]
\includegraphics[width=8cm]{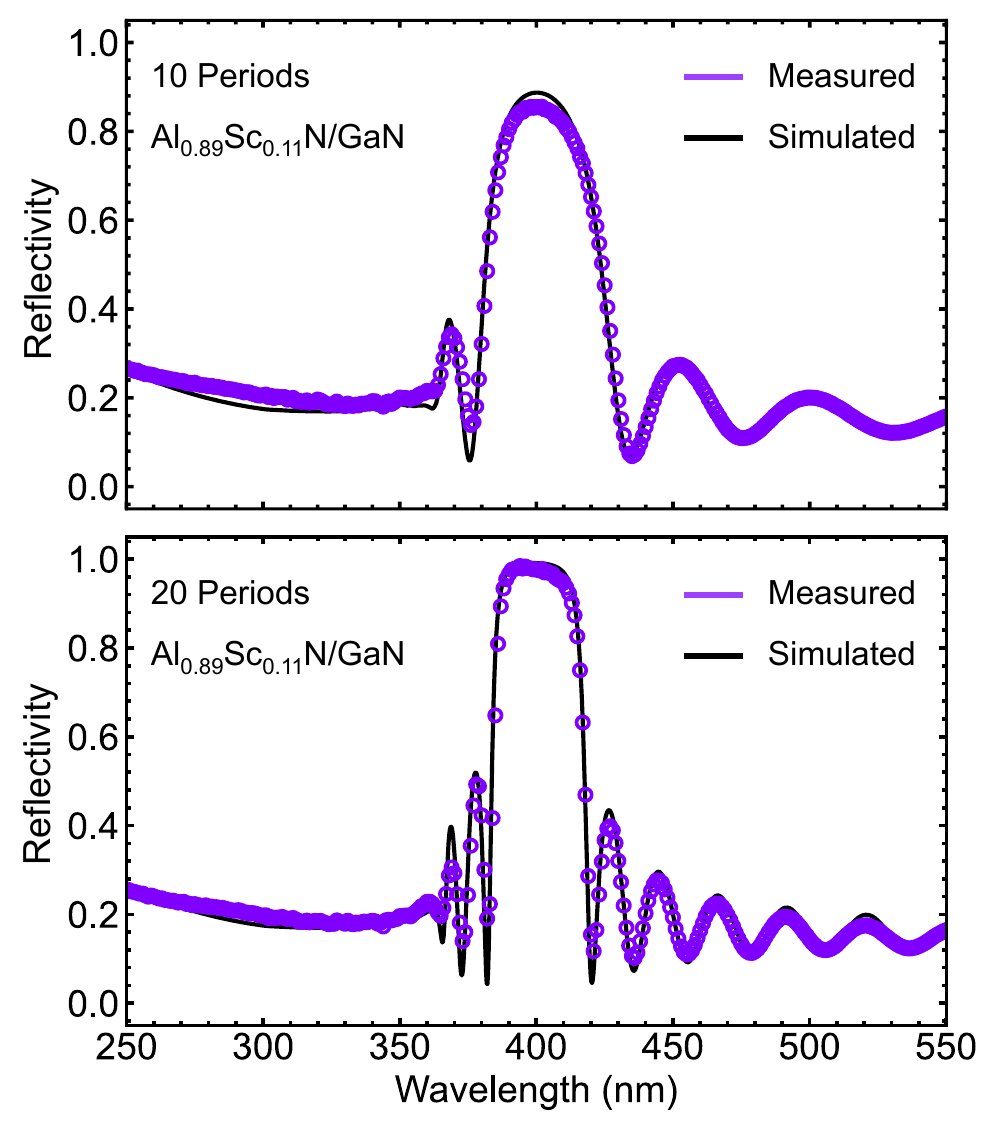}
\caption{Measured (spectrophotometry) and simulated (TMM) reflectivity spectra of a ten period (top) and twenty period (bottom) Al$_{0.89}$Sc$_{0.11}$N/GaN distributed Bragg reflector with a measured peak reflectivity of 0.86 and 0.98, respectively, for a vacuum wavelength near 400~nm.\label{fig:Figure_3}}
\end{figure}


Figure~\ref{fig:Figure_2}(c) further shows the high interface quality and precise thickness control achieved by using MBE. Sharp interfaces between AlScN and GaN layers are evidenced by the strong interference fringes in the 2$\theta$-$\omega$ scans corresponding to the ten period [Fig.~\ref{fig:Figure_2}(c)] and twenty period [see Supplementary Material Fig.~S1]. The spacing between the interference fringes matches well with a simulated multilayer structure of 10-pairs of AlScN/GaN with thicknesses of 45/40~nm per pair. Figure~\ref{fig:Figure_2}(d) shows AlScN (10$\bar{1}$5), GaN (10$\bar{1}$5) and all satellite peaks aligned vertically in the reciprocal space map, confirming that all multilayers are pseudomorphically grown on the bulk n$^{+}$GaN substrate. This would enable higher crystal quality by minimizing dislocation generation due to strain relaxation. Due to growth-to-growth flux variations, strain relaxation with 0.06\% in-plane lattice mismatch was found in AlScN layers for the twenty period sample [see Supplementary Material Fig.~S1]. By carefully tuning the Sc composition, pseudomorphic AlScN/GaN multilayer structures with more periods can be demonstrated in the future. The promising structural and surface/interface qualities indicate that lattice matched AlScN/GaN multilayer structures can serve as high quality templates and bottom reflectors for integration of active layers in vertical cavity emitters such as reported for AlInN \cite{carlinCrackfreeFullyEpitaxial2005,carlinProgressesIIInitrideDistributed2005,gacevicHighQualityInAlN2011,ikeyamaRoomtemperatureContinuouswaveOperation2016,takeuchiGaNbasedVerticalcavitySurfaceemitting2018}.


The normal incidence reflectivity spectra of the ten and twenty period DBRs near the photonic stop-band are shown in Fig.~\ref{fig:Figure_3}. As predicted from the refractive index dispersion for GaN and Al$_{0.89}$Sc$_{0.11}$N, the reflectivity spectra as simulated by the Transfer Matrix Method (TMM) match remarkably well with the experimental data for both the ten and twenty period DBRs, which are shown in Fig.~\ref{fig:Figure_3}.  This is enabled by negligible optical interface scattering losses due to the sub-nm sharp interfaces and negligible optical losses as confirmed from ellipsometry, where the ordinary optical extinction coefficient, k, of Al$_{0.89}$Sc$_{0.11}$N layers was below the detection limit ($\mathrm{k} \ll 0.001$) in the UV-A, visible and NIR regimes. The peak reflectivity was found to be 0.86 and 0.98 for the ten and twenty period DBRs, respectively, just slightly lower than the zero-loss predicted peak reflectivity values of 0.89 and 0.99. The full-width at half maximum of the photonic stop-bands are also in well agreement with the TMM simulated spectra, yielding values of 44~nm and 33~nm for the ten and twenty period DBRs, respectively.
It should be noted that below a wavelength of 365~nm, corresponding to the bandgap of GaN, the interference fringes disappear in the reflectivity spectra of Fig.~\ref{fig:Figure_3} due to the onset of interband absorption. Therefore, AlScN/GaN DBRs are limited to a photon energies lower than the bandgap of GaN. However, the ultrawide bandgap of AlScN which is larger than 5~eV for scandium contents below 25\% \cite{jinBandAlignmentScxAl12020b, casamentoEpitaxialScxAl1XN2022} make AlScN/AlGaN multilayers suitable for DBRs operating at shorter wavelengths than those limited by the bandgap of GaN, into the UV-A, UV-B and UV-C regimes.

\begin{figure}[t]
\includegraphics[width=8cm]{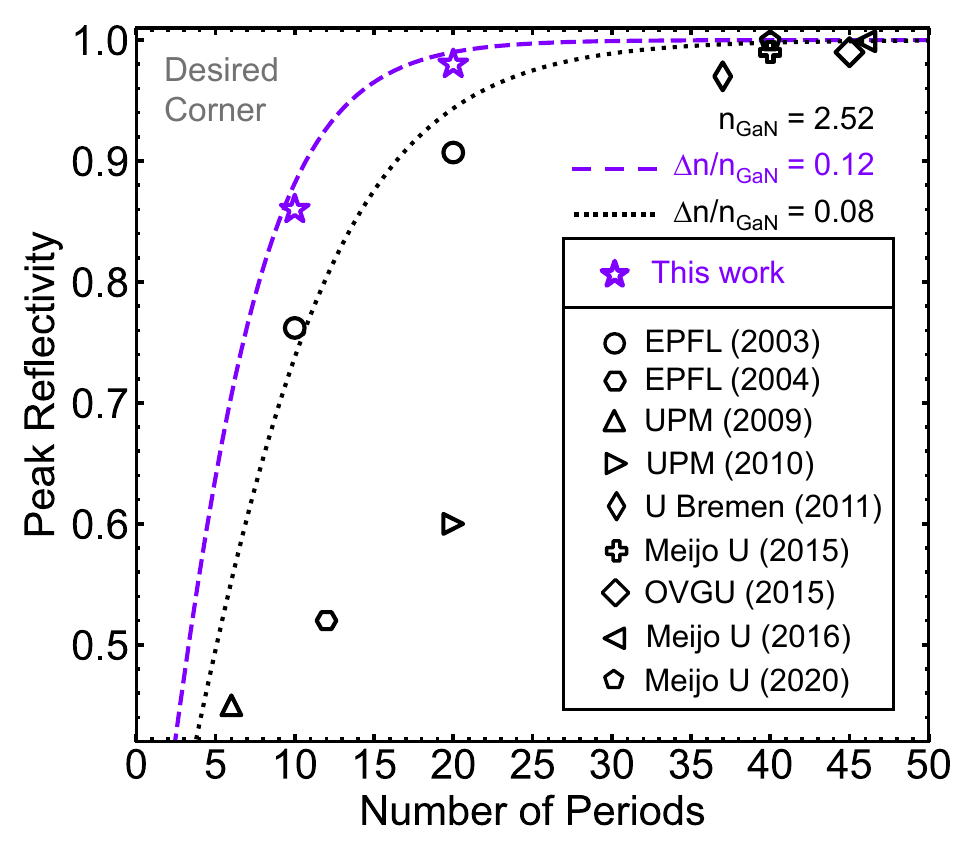}
\caption{Benchmark plot of epitaxial lattice-matched nitride based distributed Bragg reflectors on GaN showing peak reflectivity vs. number of periods in the multilayer. Reports on AlInN/GaN multilayers\cite{carlinHighqualityAlInNHigh2003a,*dorsazInGaNGaNResonantcavity2004,*gacevicGrowthCharacterizationLatticematched2009,*gacevicInAlNGaNBragg2010,*kruseGrowthCharacterizationNitridebased2011,*kozukaGrowthsAlInNSingle2014,*bergerGrowthAlInNGaN2015,*ikeyamaRoomtemperatureContinuouswaveOperation2016,*akagiHighqualityAlInNGaN2020} are show in black and AlScN/GaN based DBRs demonstrated in this report are shown by violet stars. The theoretical zero-loss approximation\cite{sheppardApproximateCalculationReflection1995a} for the reflectivity vs number of periods of GaN-based DBRs with index contrast $\mathrm{\Delta n / n_{GaN}}$ of 0.08 and 0.12 are shown in black and violet dashed curves, respectively. The ordinary refractive index of GaN, $\mathrm{n_{GaN}}$, is set to 2.52, which was the measured value at $\lambda = 400$~nm [Fig. \ref{fig:Figure_1}(e)]. \label{fig:Figure_4}}
\end{figure}


To put into perspective the advantages of using lattice-matched AlScN/GaN multilayer reflectors, we compare them with the extensively studied AlInN/GaN platform. Specifically, Figure~\ref{fig:Figure_4} shows a benchmark plot of epitaxial lattice-matched nitride based DBRs on GaN showing peak reflectivity vs number of periods in the multilayer. Reports on AlInN/GaN multilayers \cite{carlinHighqualityAlInNHigh2003a,*dorsazInGaNGaNResonantcavity2004,*gacevicGrowthCharacterizationLatticematched2009,*gacevicInAlNGaNBragg2010,*kruseGrowthCharacterizationNitridebased2011,*kozukaGrowthsAlInNSingle2014,*bergerGrowthAlInNGaN2015,*ikeyamaRoomtemperatureContinuouswaveOperation2016,*akagiHighqualityAlInNGaN2020} ranging from 400-560~nm are shown in black and AlScN/GaN based DBRs demonstrated in this report are shown by violet stars. As predicted by the lower refractive index mismatch for the AlInN/GaN platform with an index contrast of $\mathrm{\Delta n/n_{GaN}}$~=~0.08 (high estimate), AlScN/GaN outperforms AlInN/GaN due to its measured index contrast of $\mathrm{\Delta n/n_{GaN}}$~=~0.12. This is in accordance with the theoretical zero-loss approximation \cite{sheppardApproximateCalculationReflection1995a} for the reflectivity vs number of periods of GaN-based DBRs with $\mathrm{\Delta n / n_{GaN}}$ of 0.08 and 0.12 which are shown in black and violet dashed curves, respectively [Fig.~\ref{fig:Figure_4}]. The ordinary refractive index of GaN, $\mathrm{n_{GaN}}$, is set to 2.52, which is the measured value at $\lambda = 400$~nm [Fig.~\ref{fig:Figure_1}(e)]. For example, for a target peak reflectance of 0.8, a total amount of 8 periods are required for AlScN/GaN, whereas it requires 12 periods for AlInN/GaN. For a peak reflectance of 0.999, one would need 29~periods of AlScN/GaN or 45~periods of AlInN/GaN, respectively. This emphasizes that the total required material thickness is reduced substantially for a mirror with a given target reflectivity for the AlScN/GaN platform.

As a final point of discussion, we emphasize one of the reasons for the large refractive index mismatch between AlScN and GaN to be the rapid increase of the in-plane lattice parameter of AlScN as the scandium incorporation is increased, allowing for lattice-matching to GaN at high Al compositions. This is ascribed partly to the anisocrystalline alloying of rocksalt ScN with wurtzite AlN, which results in tilting of the metal-nitrogen tetrahedral bonds in the wurtzite phase, as well as the larger bond length of Sc-N as compared to Al-N \cite{casamentoEpitaxialScxAl1XN2022}. This prediction still holds true for alloying of the heavier transition metal nitrides YN or LaN with AlN. Here, the latter effect is even more significant due to the larger atomic radii of Y and La compared to Sc \cite{rowbergStructuralElectronicPolarization2021}. These considerations indicate lattice-matching to GaN at even higher Al content in AlYN and AlLaN alloys, which could result in a larger refractive index mismatch than presented here. This encourages the further exploration of transition metal nitrides for integration with group III/N optoelectronics, in particular distributed Bragg reflectors.

In summary, we have demonstrated the growth and characterization of lattice-matched Al$_{0.89}$Sc$_{0.11}$N/GaN distributed Bragg reflectors, achieving a peak reflectivity of 0.98 at a vacuum wavelength of 400 nm for a twenty period multilayer. The possibility of lattice-matching and the large refractive index contrast of $\mathrm{\Delta n/n_{GaN}}$~=~0.12 at this condition allow for higher structural quality than lattice mismatched DBRs such as those based on AlN/GaN and require fewer periods than lattice-matched AlInN/GaN DBRs. These advantages encourage the integration of transition metal nitrides with the existing III/N optoelectronic ecosystem. Such multilayer and superlattice structures are also of high interest for multichannel electronic device applications and for intersubband devices such as quantum cascade lasers.

See the Supplementary Material for structural characterizations of the twenty period DBR sample.

\section*{Acknowledgements}
Len van Deurzen and Thai-Son Nguyen contributed equally to this work. The authors thank Dr. Nina Hong and J.A. Woollam Company for assistance with accurate determination of the refractive index dispersion of the AlScN and GaN films. This work was partially supported by the Cornell Center for Materials Research with funding from the NSF MRSEC program (Grant No. DMR-1719875), as well as by the Army Research Office (Grant No. W911NF2220177).

\section*{Author Declarations}
\subsection*{Conflict of Interest}
The authors have no conflicts to disclose.
\section*{Data Availability}

The data that support the findings of this study are available from the corresponding author upon reasonable request.


\bibliography{ms}

\end{document}


\graphicspath{{./figures/}}

\title{\large{Supplementary Information - Epitaxial lattice-matched Al$_{0.89}$Sc$_{0.11}$N/GaN distributed Bragg reflectors}}

\author{L.~van~Deurzen}
\email[Electronic mail: ]{lhv9@cornell.edu}
\affiliation{School of Applied and Engineering Physics, Cornell University, Ithaca, New York 14853, USA}
\author{T.-S.~Nguyen}
\email[Electronic mail: ]{tn354@cornell.edu}
\affiliation{Department of Materials Science and Engineering, Cornell University, Ithaca, New York 14853, USA}
\author{J.~Casamento}
\affiliation{Department of Materials Science and Engineering, Cornell University, Ithaca, New York 14853, USA}
\author{H.G.~Xing}
\affiliation{Department of Materials Science and Engineering, Cornell University, Ithaca, New York 14853, USA}
\affiliation{Department of Electrical and Computer Engineering, Cornell University, Ithaca, New York 14853, USA}
\affiliation{Kavli Institute at Cornell for Nanoscale Science, Cornell University, Ithaca, New York 14853, USA}
\author{D.~Jena}
\affiliation{School of Applied and Engineering Physics, Cornell University, Ithaca, New York 14853, USA}
\affiliation{Department of Materials Science and Engineering, Cornell University, Ithaca, New York 14853, USA}
\affiliation{Department of Electrical and Computer Engineering, Cornell University, Ithaca, New York 14853, USA}
\affiliation{Kavli Institute at Cornell for Nanoscale Science, Cornell University, Ithaca, New York 14853, USA}

\begin{abstract}
The supplementary material includes:\\
Figure~S1
\end{abstract}

\maketitle

\noindent\rule{17.5cm}{0.4pt}

\section*{Supplementary Information}

\begin{figure}[H]\includegraphics[width=\textwidth]{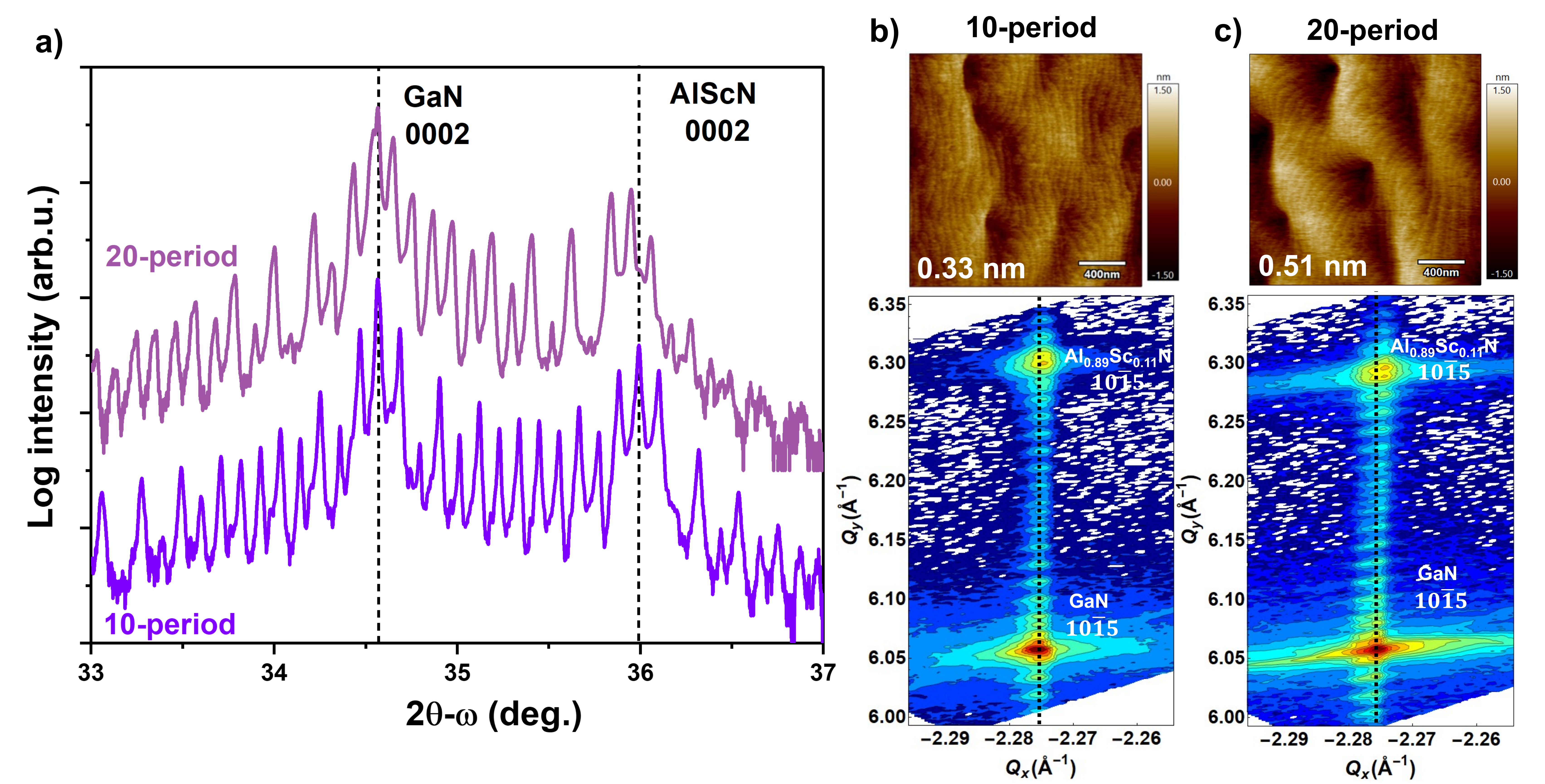}
\caption{(a) 2$\theta$-$\omega$ XRD scans of the 10-period and 20-period samples on bulk n$^{+}$GaN substrate. The interference fringe spacing suggests similar layer thicknesses were achieved in both multilayer structures. A shift in the AlScN (002) peak suggests slight variation in the Sc composition between the ten period and twenty period samples. (b, c) 2x2 $\mu$m$^2$ AFM and (10$\bar{1}$5) reflection RSM scans of the ten period and twenty period AlScN/GaN 45/40~nm multilayer structures. While the rms roughness increases from 0.33~nm for the ten period sample to 0.51~nm for the twenty period sample, clear atomic steps enabled by step-flow growth mode are maintained. The in-plane lattice mismatch between Al$_{0.89}$Sc$_{0.11}$N and GaN is 0.02\%, likely due to atomic flux drifting away from the lattice matched composition during growth. A slightly larger in-plane lattice mismatch of 0.06\% is found in the 20-period sample. By enhancing the flux control for longer growths, fully lattice-matched AlScN/GaN multilayer structures can be achieved.} 
\label{Figure_S1}
\end{figure}


\renewcommand\refname{}
\bibliography{supplement}